\documentclass[aps,twocolumn,showpacs,preprintnumbers,amsmath,amssymb]{revtex4}

\usepackage{graphicx}% Include figure files
\usepackage{dcolumn}% Align table columns on decimal point
\usepackage{bm}% bold math
\usepackage{dsfont}
\usepackage{mathrsfs}
\usepackage{amsmath}				% AMS-LaTeX-Erweiterungen
\usepackage{epstopdf}

\usepackage[bookmarks,
                 bookmarksopen = true,
                 bookmarksnumbered = true,
                 linktocpage,
                 colorlinks = true,
                 linkcolor = blue,
                 urlcolor  = blue,
                 citecolor = blue,
                 anchorcolor = green,
                 hyperindex = true,
                 hyperfigures]
		{hyperref}

\usepackage{color}

\newcommand{\vecperp}[2] {\mathbf{#1}_{\perp}^{#2}}
\newcommand{\myvec}[1] {\mathbf{#1}}
\newcommand{\Dmeson}{D^{-}}
\newcommand{\pimeson}{\pi^{-}}
\newcommand{\mylambda}{\Lambda_{c}^{+}}
\newcommand{\myprime}[2]{#1^{\prime #2}}
\newcommand{\process}{\pimeson ~p \rightarrow \Dmeson ~\mylambda}
\newcommand{\mybar}[2]{\bar{#1}^{#2}}
\newcommand{\be}{\begin{equation}}
\newcommand{\ee}{\end{equation}}
\newcommand{\myref}[1]{\( (\ref{#1}) \)}
\def\bra#1{\mathinner{\langle{#1}|}}
\def\ket#1{\mathinner{|{#1}\rangle}}
\def\braket#1{\mathinner{\langle{#1}\rangle}}
\newcommand{\gev}{~\mbox{GeV}^2}

\begin{document}
\title{$ \mathbf{ \process }$ within the Generalized Parton Picture}
\author{
Stefan Kofler$^{\,a}$, Peter Kroll$^{\,b}$,
Wolfgang Schweiger $^{\,a}$}
\affiliation{
\vspace{3mm}
$^a\,$Institut f\"ur Physik, Universit\"at Graz, 8010 Graz, Austria \\
$^b\,$Fachbereich Physik, Bergische Universit\"at Wuppertal, 42097 Wuppertal, Germany
}

\date{\today}

\begin{abstract}
\vspace{3mm}
We investigate the reaction \( \process \) within the generalized parton picture. The process
is described by a handbag-type mechanism with the charm-quark mass acting as the hard scale.
As in the case of preceding work on $\bar{p} ~p \rightarrow \bar{\Lambda}^-_c ~\mylambda$ we argue that the process amplitude factorizes into one for the perturbatively calculable partonic subprocess $\bar{u}~u\rightarrow \bar{c}~c$ and hadronic matrix elements that can be parameterized in terms of generalized parton distributions.
Modeling the generalized parton distributions by overlaps of (valence-quark) light-cone wave functions for the hadrons involved, we obtain numerical results for unpolarized differential and integrated cross sections as well as spin observables.
Our approach works well above the production threshold ($s \gtrsim 20$~GeV$^2$) in the forward hemisphere and predicts unpolarized cross sections of the order of nb, a finding that could be of interest in view of plans to measure $\process$ at J-PARC.
\end{abstract}


\maketitle

\section{Introduction}\label{sec_intro}
Hard exclusive processes have attracted much attention in recent years by both, theoreticians and experimentalists.
Above all the deeply virtual reactions as leptoproduction of mesons and photons have been theoretically studied and measured in great detail.
This interest is based on the asymptotic factorization theorems which purport that the process amplitudes can be represented as convolutions
of perturbatively calculable partonic subprocess amplitudes with generalized parton distributions (GPDs).
This, so-called, handbag approach is quite successful in describing the deeply virtual processes
qualitatively as well as quantitatively.
An alternative class of hard exclusive processes is characterized by large Mandelstam $-t$ (and $-u$) providing the hard scale.
For this class the amplitudes factorize in a product of subprocess amplitudes and form factors representing moments of GPDs
Examples of such processes are wide-angle real Compton scattering  or time-like reactions as, e.g., two-photon annihilations into pairs of hadrons.
Also the time-reversed process proton-antiproton annihilation into two photons (or photon and meson) belong to this class.
Again the handbag approach works very well.
A particular outstanding example is real Compton scattering.
A GPD analysis of the nucleon form factors provided also results for the Compton form factors.
Hence, Compton scattering in the wide-angle region can be evaluated free of parameters.
The results are found to be in fair agreement with experiment.
New measurements performed at the upgraded Jlab will provide another crucial test for the quality of these results.
Future precise data from BELLE and FAIR may further probe the predictions for the time-like processes.

A third class of hard exclusive processes, which are amenable to the handbag approach, is formed by reactions involving heavy hadrons.
Here the large scale is set by the heavy-quark mass and the model can be applied to the forward hemisphere and Mandelstam $s$ well above the reaction threshold.
Like for the wide-angle processes the heavy-hadron amplitudes are represented by products of subprocess amplitudes and appropriate form factors.
Till now the processes $\bar{p} p \rightarrow \bar{\Lambda}_c^- \Lambda_c^+$ \cite{schweig1},  $\bar{p} p \rightarrow  D^0 \overline{D}^0$ \cite{gor} and
$\gamma p\to \overline{D}^0 \Lambda_c^+$ \cite{kofler} have been investigated.
An experimental verification of the derived results is still pending, there are no data as yet.
A considerable improvement of the experimental situation for hadron pair production is to be expected from the upcoming $\bar{\mathrm P}$ANDA detector at FAIR.
The photoproduction of \( \overline{D}^0 \)-mesons could be, tentatively, measured at the upgraded JLab.

In this work we are going to investigate a pion-induced process, namely $\pi p\to D^-\Lambda_c^+$,
within the handbag approach.
After a few kinematical preliminaries in Sec.~\ref{sec_kin}, we sketch in Sec.~\ref{sec_mech} the handbag approach to the process of interest.
In Sec.~\ref{sec_results} we present numerical results for cross sections and polarizations.
The paper ends with a summary and our conclusions, Sec.~\ref{sec_sum}.

\section{Hadron Kinematics}\label{sec_kin}
The momenta, light-cone (LC) helicities and masses of the incoming proton and \( \pimeson\) are denoted
by \( p,~\mu,~m_{p} \) and \(q,~m_{\pi} \), those of the outgoing \( \mylambda \) and \( \Dmeson \) by
\( \myprime{p}{},~\myprime{\mu}{},~M_{\Lambda_c} \) and \( \myprime{q}{},~M_D\), respectively.
We consider the reaction in a symmetric center-of-momentum system (CMS) which has the \( z \)-axis aligned along the three-vector part, \( \bar{\mathbf{p}} \), of the average momentum \( \bar{p} \equiv \frac{1}{2} \left(p + p^{\prime}\right)\).
This reference frame is chosen such that the transverse component of the momentum transfer \( \Delta \equiv \left(p^\prime- p  \right) = \left( q - q^\prime \right) \) is symmetrically shared between the particles.
Introducing the skewness parameter
\be
\label{eq_skewness}
\xi \equiv \frac{p^{+} - \myprime{p}{+}}{p^{+} + \myprime{p}{+}} = -\frac{\Delta^{+}}{2 \mybar{p}{+}},
\ee
we parameterize the proton and the \( \mylambda \) momenta as follows \footnote{We use LC
coordinates and the Kogut-Soper convention~\cite{Brodsky:LightCone}, where a four-vector is then written as
\( \left[a^+,a^-, \vecperp{a}{}  \right] \) with
\( a^{\pm} \equiv \frac{1}{\sqrt{2}} \left(a^0 \pm a^3 \right) \) and \( \vecperp{a}{} \equiv (a^1, a^2)\).}:
\begin{align}
\begin{aligned}
\label{eq_mom1}
p &= \Bigg[ (1 + \xi) \mybar{p}{+}, \frac{m_{p}^{2} + \vecperp{\Delta}{2}/4}{2(1 + \xi) \bar{p}^{+}},-\frac{\vecperp{\Delta}{}}{2} \Bigg],  \\
\myprime{p}{} &= \Bigg[ (1 - \xi) \mybar{p}{+}, \frac{M_{\Lambda_c}^{2} +  \vecperp{\Delta}{2}/4}{2(1 - \xi) \mybar{p}{+}},\frac{\vecperp{\Delta}{}}{2} \hspace{0.07cm} \Bigg].
\end{aligned}
\end{align}
The \( \pimeson\)-meson and the \( \Dmeson \)-meson momenta can be written in an analogous way:
\begin{align}
\begin{aligned}
\label{eq_mom2}
q  &= \Bigg[ \frac{m_{\pi}^{2} + \vecperp{\Delta}{2}/4}{2(1 + \eta) \bar{q}^{-}},~ (1 + \eta) \mybar{q}{-}, ~~\frac{\vecperp{\Delta}{}}{2} \Bigg], \\
\myprime{q}{}  &=  \Bigg[ \frac{M_{D}^{2} + \vecperp{\Delta}{2}/4}{2(1-\eta)\mybar{q}{-}},(1-\eta)\mybar{q}{-}, -\frac{ \vecperp{\Delta}{}}{2} \Bigg],
\end{aligned}
\end{align}
\noindent with
\be
\label{eq_qbar_eta_def}
\bar{q} \equiv \frac{1}{2}(q + q^{\prime}) \hspace{0.5cm} \mbox{and} \hspace{0.5cm}  \eta \equiv \frac{q^{-} - \myprime{q}{-}}{q^{-} + \myprime{q}{-}} = \frac{\Delta^{-}}{2\bar{q}^{-}}.
\ee
$\bar{q}^-$ and $\eta$ are determined by our CMS kinematics ($\myvec{p}+\myvec{q}=0$, $\myvec{p}^{\,\prime}+\myvec{q}^{\,\prime}=0$). The relation between $\xi$ and $\eta$ is most easily obtained from $\Delta^\pm=(p^\prime-p)^\pm=(q-q^\prime)^\pm$.

\section{Double-Handbag Mechanism, Factorization and GPDs}\label{sec_mech}
\begin{figure}[h!]
\centering
\includegraphics[width=.45\textwidth]{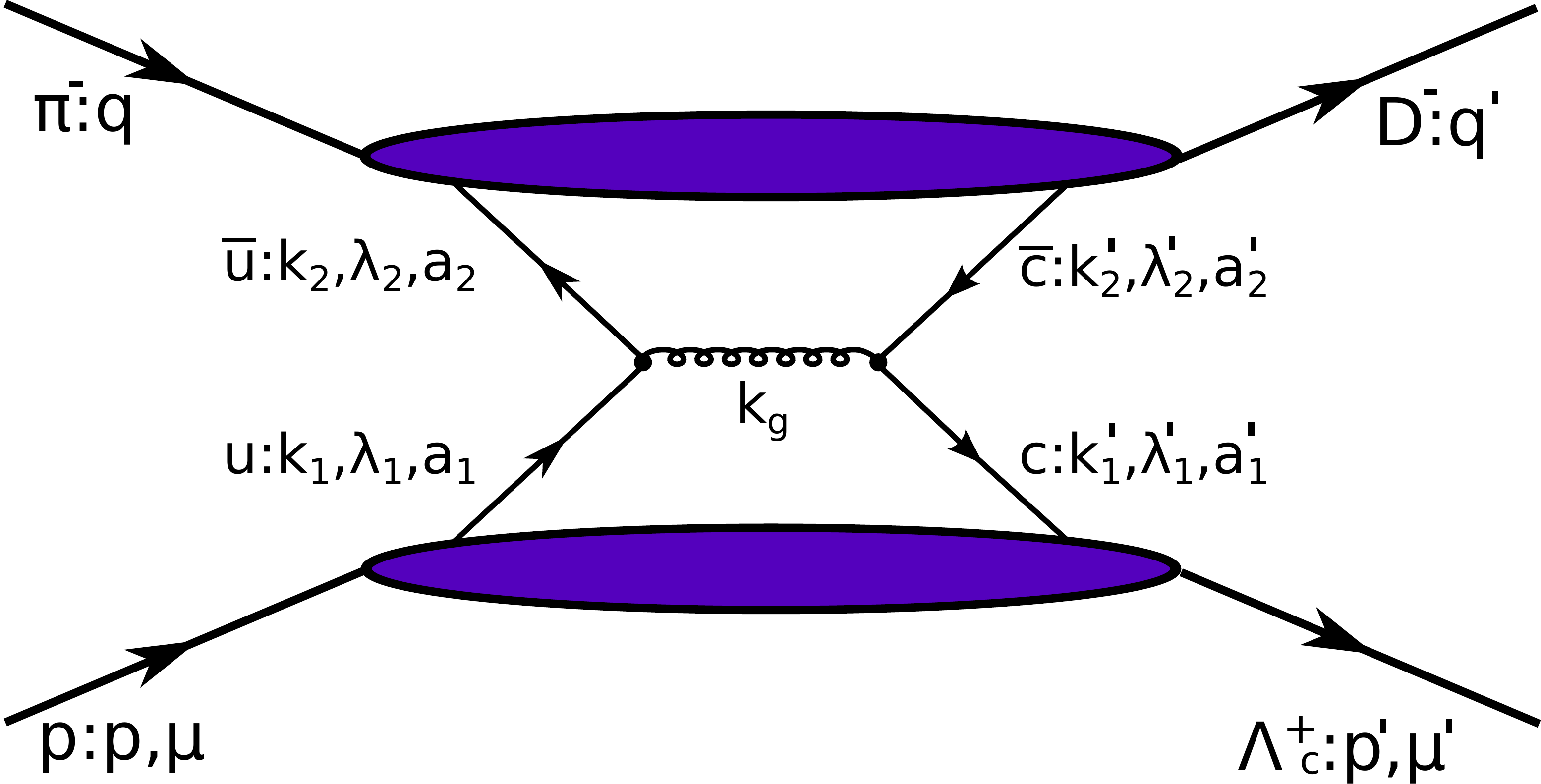}
\caption{The double-handbag contribution to the process \( \process \) (in the DGLAP region). The momenta,
LC helicities and colors of the quarks are specified.}
\label{fig_handbag}
\end{figure}

As in Ref.~\cite{schweig1} we argue that intrinsic (non-perturbative) charm of the proton can be neglected and the mechanism which dominates $\process$ well above the kinematical threshold \big(\((M_{\Lambda_c} + M_D)^{2} \approx 17.27 \gev\)\big) and in the forward hemisphere is the one depicted in Fig.~\ref{fig_handbag}.
The reasoning for factorization of this handbag-type mechanism goes along the same lines as in Ref.~\cite{schweig1}.
One has to assume that the parton virtualities and (intrinsic) transverse momenta are restricted by a typical hadronic scale of the order of \( 1\)~GeV and, in addition, that the $p\rightarrow \Lambda_c$ ($\pi \rightarrow D$)  GPDs exhibit a pronounced peak at a large value of $\bar{x}_1$ ($\bar{x}_2$) close to the ratio of charm-quark and charmed-hadron masses $\bar{x}_{10}=m_c/M_{\Lambda_c} \approx 0.56$ ($\bar{x}_{20} = m_c/M_D \approx 0.68$).
Such a behavior parallels the theoretical expected and experimentally confirmed property of heavy-quark fragmentation functions, in particular for $c\to\Lambda_c^+$~\cite{kniehl-kramer} and is also analogous to the behavior of heavy-hadron distribution amplitudes (DAs)\cite{KK,BB}.
Under these assumptions, with $m_c$ taken as the hard scale, the hadronic amplitude $\mathscr{M}$ is seen to factorize in a hard partonic scattering kernel $\tilde{H}$ and soft hadronic matrix elements which describe the $p\rightarrow \Lambda_c$ and $\pi \rightarrow D$ transitions by emission and absorption of  soft (anti)quarks.
These quarks participate in the partonic subprocess \(\bar{u}\, u   \rightarrow \bar{c}\,  c \), are approximately on-mass-shell and collinear with their parent hadron.
The resulting formal expression for the process amplitude (for details of the derivation, see Refs.~\cite{schweig1,wideangle}) reads \footnote{In Eq.~(\ref{eq_amp_simp}) we have tacitly assumed that we are working in the LC gauge  \( A^+ = 0\).}:
\begin{widetext}
\begin{align}
\begin{aligned}
\label{eq_amp_simp}
    \mathscr{M}_{\mu^{\prime}, \mu} &= \sum_{a_{i}^{(\prime)},\alpha_{i}^{(\prime)}} \int \mathrm{d} \bar{x}_{1} \,\bar{p}^{+}  \int \frac{\mathrm{d} z^{-}_{1}}{(2 \pi)} e^{i \bar{x}_{1} \bar{p}^{+} z^{-}_{1}} \int \mathrm{d} \bar{x}_{2} \, \bar{q}^{-} \int \frac{\mathrm{d} z_{2}^{+} }{(2 \pi)} e^{i \bar{x}_{2} \bar{q}^{-} z_{2}^{+} } \\
    &\times \braket{\Lambda_{c}^{+}:p^{\prime}, \mu^{\prime} | ~\overline{\Psi}^{c}_{a_1^\prime,\alpha_1^\prime} \left( - \frac{z^{-}_{1}}{2} \right) \Psi^{u}_{a_1,\alpha_1} \left( \frac{z^{-}_{1}}{2} \right)~|p: p, \mu}~ \widetilde{H}_{a_{i}^{(\prime)},\alpha_{i}^{(\prime)}} \left(\bar{x}_{1} \bar{p}^{+},\bar{x}_{2} \bar{q}^{-}\right) \\
  &\times \braket{\Dmeson: q^{\prime}|~\overline{\Psi}^{u}_{a_{2},\alpha_{2}} \left(  \frac{z^{+}_{2}}{2} \right) \Psi^{c}_{a_{2}^{\prime},\alpha_{2}^{\prime}}\left( - \frac{z^{+}_{2}}{2} \right)~ |\pimeson:q},
\end{aligned}
\end{align}
\end{widetext}
with \( a_{i}^{(\prime)} \) and \( \alpha_{i}^{(\prime)} \) denoting color and Dirac indices and the average momentum fractions of the active (anti)quarks
\be
\label{eq_average_mom_fracs}
\bar{x}_{1} \equiv \frac{k_{1}^{+} + k_{1}^{\prime + }}{p^+ + p^{\prime +} } = \frac{\bar{k}_1^+}{\bar{p}^{+}} \hspace{0.5cm} \mbox{and} \hspace{0.5cm} \bar{x}_{2} \equiv \frac{k_{2}^{-} + k_{2}^{\prime - }}{q^- + q^{\prime -} } = \frac{\bar{k}_2^-}{\bar{q}^{-}}.
\ee
For the assignment of (anti)quark momenta, helicities as well as color and Dirac indices, see Fig.~\ref{fig_handbag}.
The line of arguments leading to Eq.~(\ref{eq_amp_simp}) puts an upper bound on $\vecperp{\Delta}{2}$ which restricts the validity of our approach to a particular angular range around the forward direction.
For the energies we are interested in this angular range is, however, sufficiently large to obtain reasonable estimates for integrated cross sections.
Using projection techniques as in Ref.~\cite{schweig1} we pick out the ``leading twist'' contributions from the bilocal quark-field operator product \( \overline{\Psi}^{c}(-z_1^-/2)  \Psi^{u}(z_1^-/2): \)
\be
\bra{\mylambda}\overline{\Psi}^{c}  \Psi^{u} \ket{p} : \bra{\mylambda} \overline{\Psi}^{c} \big \{\gamma^{+}, \gamma^{+}\gamma_5 , i\sigma^{+j} \big \} \Psi^{u} \ket{p} \label{eq_operator_twist1}
\ee
and from \( \overline{\Psi}^u(z_2^+/2) \Psi^c(-z_2^+/2):\)
\be
 \bra{\Dmeson}\overline{\Psi}^u \Psi^c\ket{\pimeson} :\bra{\Dmeson}  \overline{\Psi}^u \big \{ \gamma^{-}, \gamma^{-} \gamma_5, i\sigma^{-j} \big \} \Psi^c \ket{\pimeson}, \label{eq_operator_twist2}
 \ee
respectively (\( \sigma^{\pm j} = i \gamma^{\pm} \gamma^j \) with \( j =1,2\) labeling transverse components).
The three Dirac structures showing up in Eqs.~\myref{eq_operator_twist1} and \myref{eq_operator_twist2} can be considered as $+$ or $-$ components of (bilocal) vector, pseudovector and tensor currents, respectively.
These currents are then Fourier transformed (with respect to $z_1^-$ or $z_2^+$, respectively) and decomposed into appropriate hadronic covariants.
The  coefficients in front of these covariants  are the quantities which are usually understood as GPDs.
For the \( p \rightarrow \Lambda_c^+ \) transition this kind of analysis leads to 8 GPDs, as explained in some detail in Ref.~\cite{schweig1}.
Matters become much simpler for the pseudoscalar to pseudoscalar \( \Dmeson \rightarrow \pimeson \) transition.
Due to parity invariance the matrix elements \( \bra{\Dmeson} \overline{\Psi}^u \gamma^{-} \gamma_5 \Psi^c \ket{\pimeson} \) vanish and the covariant decomposition of the remaining vector and tensor currents gives rise to two \(\pimeson \rightarrow \Dmeson \) transition GPDs,  \( H^{\overline{cu}}_{\pi D} \) and \( E_{T \pi D}^{\overline{cu}} \), which are defined by~\cite{Brommel} \footnote{This definition resembles also the one for the matrix elements $\bra{\mylambda} \overline{\Psi}^{c} \gamma^{+} \Psi^{u}\ket{p}$ and $\bra{\mylambda} \overline{\Psi}^{c} i\sigma^{+j} \Psi^{u} \ket{p}$ introduced in Ref.~\cite{schweig1}.}:
\begin{widetext}
\be
\begin{split}
\label{eq_mesonic_GPDs}
\bar{q}^{-} \int \frac{d z^{ +}_{2}}{2\pi} e^{i \bar{x}_{2} \bar{q}^{-}z^{+}_{2}}
                \bra{\Dmeson:q^\prime }   \overline{\Psi}^{u}(z^{+}_{2} /2) \Big \{ \gamma^- , i \sigma^{-j} \Big \} \Psi^{c}(-z^{+}_{2}/2) \ket{\pimeson:q} \\
               = \Big \{ 2 \mybar{q}{-}\, H^{\overline{cu}}_{\pi D} (\bar{x}_{2},\eta,t), \frac{\mybar{q}{-} \Delta^{j} -  \Delta^{-} \mybar{q}{j} }{m_\pi + M_D} E_{T \pi D}^{\overline{cu}} (\bar{x}_{2},\eta,t)  \Big \}.
\end{split}
\ee
\end{widetext}
These GPDs are functions of the average momentum fraction \(\bar{x}_{2} \), the skewness parameter \( \eta \) and the Mandelstam variable \( t=\Delta^2 \).

Having expressed the soft hadronic matrix elements in Eq.~(\ref{eq_amp_simp}) in terms of generalized parton distributions one ends up with an integral in which these parton distributions, multiplied with the hard partonic scattering amplitude $H_{\lambda_1^\prime \lambda_2^\prime ,\lambda_1 \lambda_2}\left(\bar{x}_{1} \bar{p}^{+},\bar{x}_{2} \bar{q}^{-}\right)$, are integrated over $\bar{x}_1$ and $\bar{x}_2$.
The requirement for Mandelstam $s$ to be large enough to produce the $c\bar c$ pair puts some kinematical constraints on $\bar{x}_1$ and $\bar{x}_2$. For $s$ well above the production threshold ($s \gtrsim 20$~GeV$^2$) and in the forward-scattering hemisphere it can be checked numerically that $\bar{x}_1 > \xi$ and $\bar{x}_2 > \eta$.
This means that the ERBL region ($\bar{x}_1 <  \xi$, $\bar{x}_2 < \eta$) does not contribute in our case.
The supposition that the  \( p \rightarrow \Lambda_c^+ \) and \( \Dmeson \rightarrow \pimeson \) GPDs are strongly peaked at $\bar{x}_{10}$ and $\bar{x}_{20}$, respectively, leads to a further simplification of the $\process$ amplitude.
The major contributions to the $\bar{x}_1$ and $\bar{x}_2$ integrals will then come from $\bar{x}_1\approx \bar{x}_{10}$ and $\bar{x}_2\approx \bar{x}_{20}$.
One can thus replace the hard partonic scattering amplitude by its value at the peak position, $H_{\lambda_1^\prime \lambda_2^\prime ,\lambda_1 \lambda_2}\left(\bar{x}_{10} \bar{p}^{+},\bar{x}_{20}\bar{q}^{-}\right)$ and take it out of the integral.
What one is left with are separate integrals over the GPDs which may be interpreted as generalized \( p \rightarrow \Lambda_c^+ \)  and \( \Dmeson \rightarrow \pimeson \) transition form factors.
In the formal limit of $m_c\to\infty$ $\bar{x}_{10}$ and $\bar{x}_{20}$ tend to 1 according to the heavy-quark effective theory~\cite{IW1991}.
This makes it obvious that our approach can be viewed as a variant of the familiar Feynman mechanism.
With this \lq\lq peaking approximation\rq\rq\ our final expressions for the $\process$ amplitudes become:
\begin{align}
\label{eq_procamp}
 &\mathscr{M}_{+,+} =\mathscr{M}_{-,-} =\frac{1}{4}  \sqrt{1 - \xi^2} \, H_{+-,+-} \, R_V \, G\, , \nonumber \\
&\mathscr{M}_{+,-} =  -\mathscr{M}_{-,+}  =\frac{1}{4}   \sqrt{1 - \xi^2} \,  H_{++,-+}\, S_T \, G,
\end{align}
\noindent with the \(\pimeson \rightarrow \Dmeson\) transition form factor
\be\label{eq_piDff}
G(\eta,t) = \int_{\eta}^{1} \frac{\mathrm{d} \bar{x}_2 }{\sqrt{\bar{x}^{2}_2 - \eta^2}}
H^{\overline{cu}}_{\pi D}(\bar{x}_2,\eta,t) \, .
\ee
In Eqs.~(\ref{eq_procamp}) we have restricted ourselves to the two most important $p\to\Lambda_c$ GPDs, $H^{cu}_{p\Lambda_c}$ and $H^{cu}_{T p\Lambda_c}$, leading to the respective form factors $R_V$ and $S_T$, defined analogously to Eq.~(\ref{eq_piDff}).
The underlying assumption is that those GPDs (and corresponding form factors) which involve non-zero orbital angular momentum of the (anti)quarks that make up the hadrons are suppressed.
This leads also to omission of \( E_{T \pi D}^{\overline{cu}}\).

The $H_{\lambda_1^\prime \lambda_2^\prime ,\lambda_1 \lambda_2}$ are LC helicity amplitudes for $u\bar{u}\rightarrow c\bar{c}$ via one-gluon exchange \footnote{For the definition and normalization of the LC-helicity spinors, see Ref.~\cite{gor}.}. Naive application of the collinear approximation gives
(minus signs for primed momenta) $k_1^{(\prime)}=(\bar{x}_{10}\pm\xi) p^{(\prime)}/(1\pm\xi)$ and $k_2^{(\prime)}=(\bar{x}_{20} \pm\eta) q^{(\prime)}/(1\pm\eta)$ for the parton momenta ($m_p$ and $m_\pi$ are usually neglected). In order to match the subprocess kinematics (charm-quark mass $m_c$) with the one on the hadronic level (hadron masses $M_{\Lambda_c}\neq M_D$) some further approximations are required. As one can easily verify $k_1+k_2\neq k_1^\prime + k_2^\prime$, i.e. momentum conservation does not hold on the partonic level, in general. There are only two special cases in which momentum conservation is recovered. The first case is $\bar{x}_{10},\bar{x}_{20} \rightarrow 1$, which one would obtain in the heavy-quark limit ($M_{\Lambda_c}=M_D=m_c\rightarrow\infty$). The second case is $\bar{x}_{10}=\bar{x}_{20}$ finite, but $\xi=\eta\simeq 0$, which holds for finite charm-quark mass in the limit of large (hadronic) Mandelstam $s$. In these two limiting cases the partonic amplitudes become formally the same if expressed in terms of the hadronic momentum components $p^{+(\prime)}$, $q^{-(\prime)}$, $\mathbf{\Delta}_\perp$ and Mandelstam $s$. They only differ in the argument of the strong coupling $\alpha_s$ which is Mandelstam $s$ in the first case and $(\bar{x}_{10} \bar{x}_{20}\, s)$ in the second one. Since we want apply our approach for physical masses of the heavy hadrons it seems more plausible to take $(\bar{x}_{10} \bar{x}_{20}\, s)$ as the scale which determines the strength of $\alpha_s$. In both cases one demands that $\bar{x}_{10}=\bar{x}_{20}$ which means that an average mass must be taken for the heavy hadrons when calculating the partonic amplitude. We take the geometric mean value $M^2=M_{\Lambda_c} M_D$. The resulting analytic expressions for $H_{+-,+-}$, $H_{+-,-+}$ and $H_{++,-+}$ are given in App.~\ref{app_part}.

 \begin{figure*}[t]
\centering
\includegraphics[width=.48\textwidth]{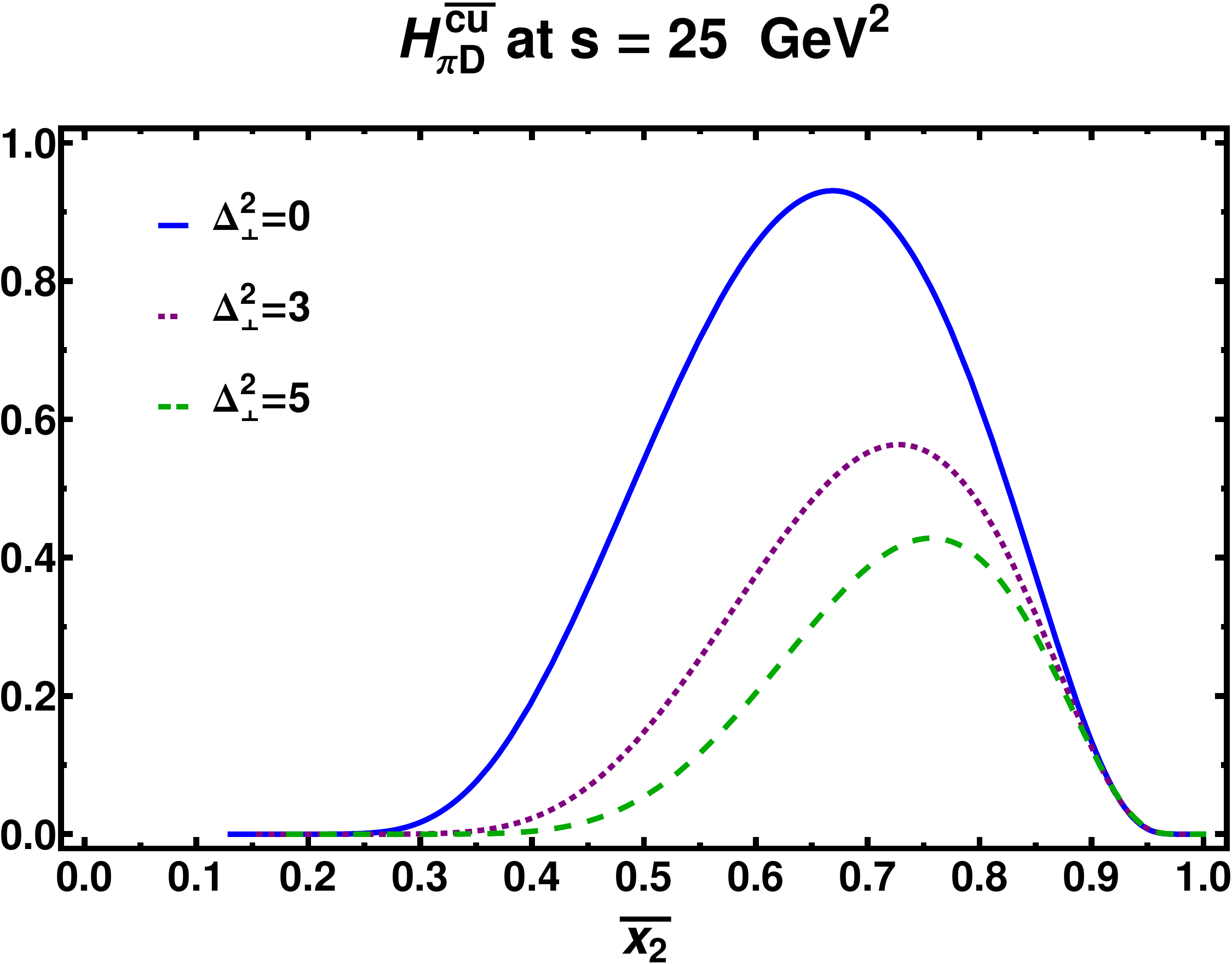}
\hspace{0.2cm}
\includegraphics[width=.48\textwidth]{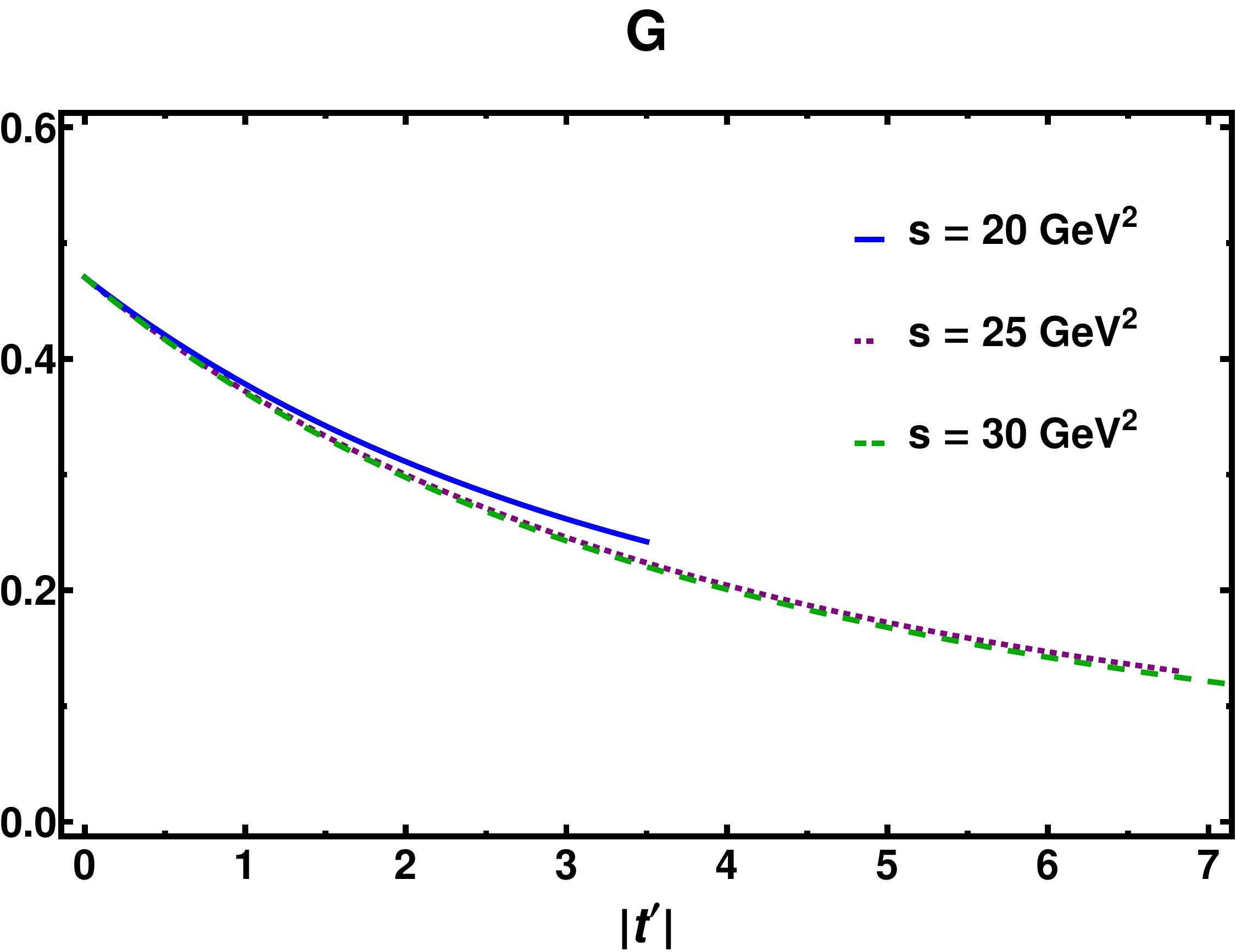}
\caption{The left plot shows the $\pimeson \rightarrow \Dmeson$ transition GPD \( {H}^{\overline{cu}}_{\pi D} \) vs. \( \bar{x}_2 \) at Mandelstam \( s = 25 \gev \) and $\mathbf{\Delta}_\perp^2=0$, $3.0$, $5.0$~GeV$^2$ (solid, dotted and dashed line), corresponding to ${ |t^\prime|=|t-t_0|}=0$, $3.31$, $5.69$~GeV$^2$ ($\eta=0.13$, $0.16$, $0.18$), for the hadron LCWFs introduced in the text (KK mass exponential). $t_0$ is the (non-vanishing) value of $t$ for forward scattering ($\mathbf{\Delta}_\perp=0$, $p^{\prime 3}\geq 0$). The right plot shows the corresponding $\pi^- \rightarrow D^-$ transition form factor $G$ as function of $|t^\prime|$ for \( s = 20,~ 25~\mbox{and}~30 \gev \) (solid, dotted and dashed line).
}
\label{fig_GPDs}
\end{figure*}
In order to make numerical predictions for $\process$ observables we need to know how the GPDs and, in particular, the form factors $R_V$, $S_T$ and $G$ look like.
This requires some modeling.
The fact that contributions from the ERBL region are suppressed for the kinematical situations we are interested in, allows for an overlap representation of the GPDs in terms of LC wave functions (LCWFs) of the valence Fock-states of $p$,  $\Lambda_c$, $\pi$ and $D$.
Proceeding along the lines of Refs.~\cite{overlap,Brodsky:2000xy} the overlap representation for the $\pi^- \rightarrow D^-$ transition GPDs is, e.g., obtained by inserting the Fourier representation of the field operators and the Fock-state decomposition of the corresponding hadron states in LC-quantum field theory into the left-hand side of Eq.~(\ref{eq_mesonic_GPDs}).
The restriction to valence (anti)quarks is supposed to be a good approximation for $\Lambda_c$ and $D^-$.
For the $p$ and the $\pi^-$ higher Fock states are most likely also important, but they do not contribute to the pertinent matrix elements with the valence Fock states of $\Lambda_c$ and $D^-$, respectively.

We take simple s-wave wave functions for the hadron ground states.
This has the consequence that \( \bra{\Dmeson} i \sigma^{-j} \ket{\pimeson} \) vanishes.
The reason is that the tensor structure requires the flip of a quark helicity which means that in at least one of the LCWFs, \( \psi_\pi \) or \( \psi_D \), the helicity of the meson is not the sum of its parton helicities so that orbital excitations of the quarks have to come into play.
For zero orbital angular momentum the \( \pimeson \rightarrow \Dmeson \) transition matrix element can thus be expressed in terms of a single GPD, namely \(  H^{\overline{cu}}_{\pi D}\).
Likewise, for pure s-wave baryon wave functions five of the eight $p\rightarrow \Lambda_c$ transition GPDs vanish and only $H^{cu}_{p\Lambda_c}$, $\tilde{H}^{cu}_{p\Lambda_c}$ and $H^{cu}_{Tp\Lambda_c}$ survive~\cite{schweig1}.
For a reasonably small probability to find the $c$ quark with helicity opposite to the one of the $\Lambda_c$ these three GPDs are approximately the same.
As already mentioned we take into account only the form factors $R_V$ and $S_T$ and adopt the numerical results for them from Ref.~\cite{schweig1} for the present calculation.
In this work the wave function suggested by Bolz and Kroll~\cite{Bolz:1996sw}, which is supported by several phenomenological applications, has been taken for the proton.
A slightly modified version of it, with an additional mass exponential that provides the expected pronounced peak at $\bar{x}_{10}$, was taken for the $\Lambda_c$~\cite{KK}.

The wave functions of the $\pimeson$ and $\Dmeson$ are parameterized in a quite analogous way.
For the \( \pimeson \) we use
\be
\label{eq_pimson_LCWF}
\psi_{\pi} \left ( \tilde{x}^\prime, \tilde{\mathbf{k}}_{\perp}^{\prime} \right)= N_\pi \, \exp \left[ \frac{-a_{\pi}^{2} \, \tilde{\mathbf{k}}_{\perp}^{\prime 2} }{\tilde{x}^\prime (1-\tilde{x}^\prime)}\right]
\ee
with the parameters $N_\pi=18.56$~GeV$^{-2}$ and $a_\pi=0.85$~GeV$^{-1}$ taken from Ref.~\cite{Feldmann}.
This wave function gives rise to the asymptotic DA $ { \phi_\pi^{asy}}(x)=6 \, x \, (1-x)$, reproduces the pion decay constant $f_\pi=0.132$~GeV and provides a valence-Fock-state probability of $P_\pi=0.25$.
Like the $\Lambda_c$ our \( \Dmeson \)-LCWF contains also an additional mass exponential (see Ref.~\cite{KK}):
\begin{align}
\begin{aligned}
\label{eq_Dmeson_LCWF}
 \psi_{D} \left( \hat{x} ^\prime, \hat{\mathbf{k}}_{\perp}^{\prime} \right)  &= N_{D} \, \exp \left[- a_{D}^{2} M_{D}^2 \frac{\left( \hat{x}^\prime-{x}^{\prime}_{0} \right)^2}{\hat{x}^\prime \left( 1- \hat{x}^\prime \right)} \right] \\
 &\times \exp \left[\frac{-a_{D}^{2} \, \hat{\mathbf{k}}_{\perp}^{ \prime 2} }{\hat{x}^\prime (1-\hat{x}^\prime )}\right].
\end{aligned}
\end{align}

\begin{figure*}[t]
\centering
\includegraphics[width=.47\textwidth]{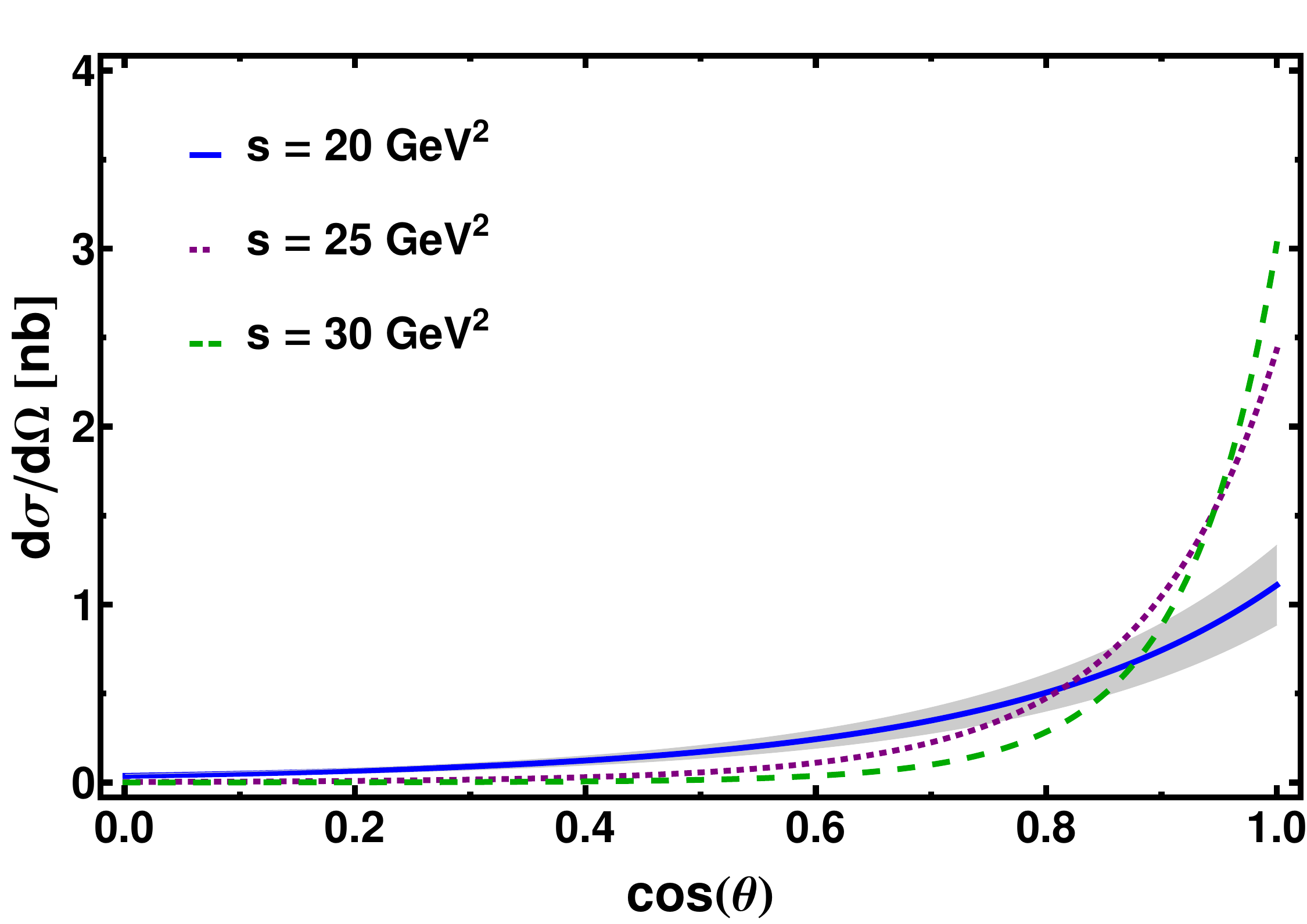}\hspace{0.5cm}
\includegraphics[width=.48\textwidth]{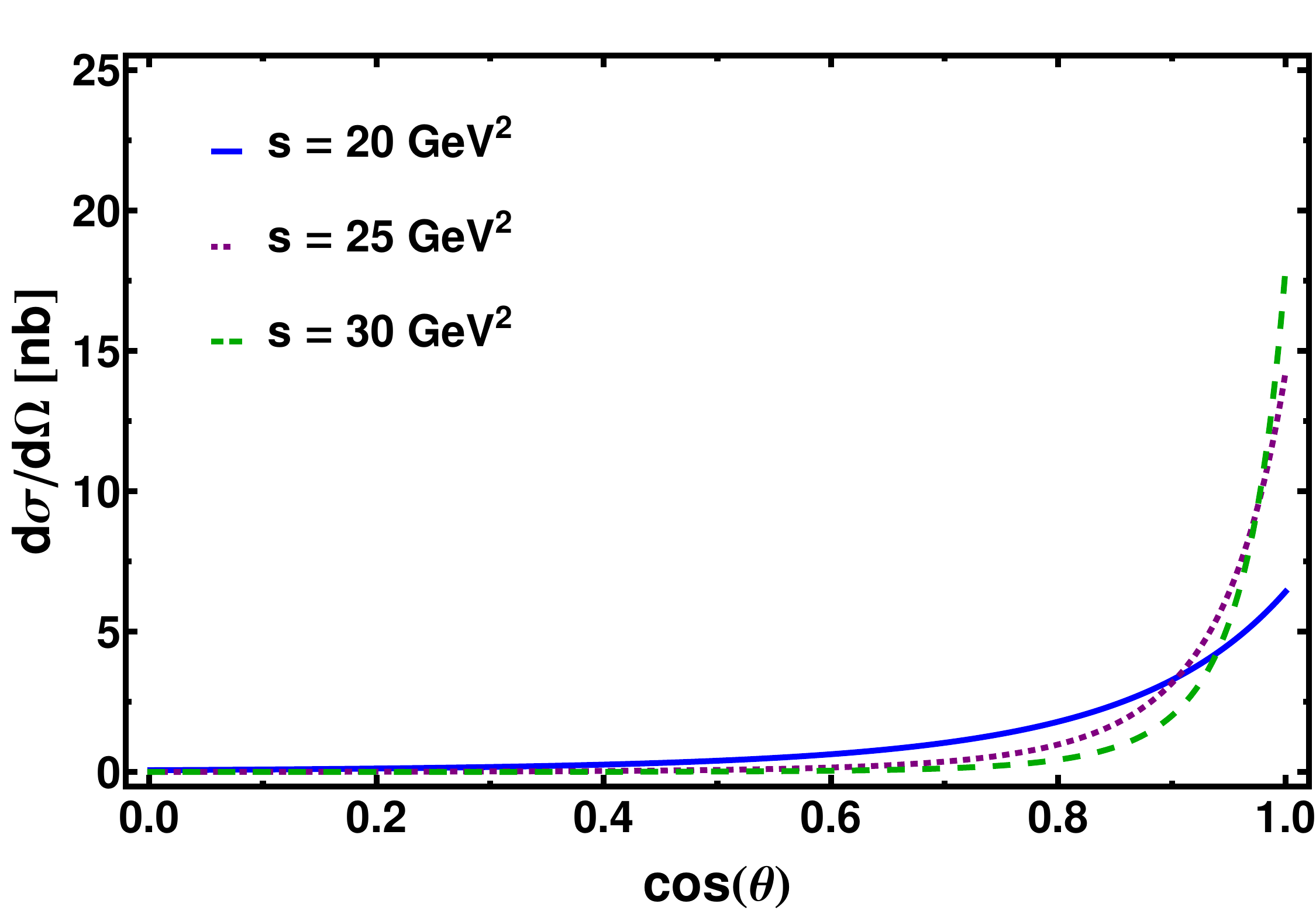}
\caption{The differential $\process$ CM cross section versus $\cos\theta$ for \( s = 20,\,25,\, 30 \gev \) (solid, dotted and dashed line). The left plot  has been obtained with the wave-function parameterizations described in the text (KK mass exponential). The effects of uncertainties in the $\Lambda_c$ and the $D^-$ wave-function parameters are indicated by the shaded band around the $s=20 \gev$ curve. Results obtained with a different analytic form of the $\Lambda_c$ and $D^-$ LCWFs  (BB mass exponential) are shown in the right plot.}
\label{fig_dcs}
\end{figure*}

The parameters $N_D=54.92$~GeV$^{-2}$ and $a_D=0.86$~GeV$^{-1}$ are chosen such that the experimental value of the D-meson decay constant $f_D=0.207$~GeV \cite{PDG} is reproduced and the valence-Fock-state probability becomes $P_D=0.9$. The mass exponential chosen here corresponds to the one for the $\Lambda_c$ which was denoted by \lq\lq KK\rq\rq\ in Ref.~\cite{schweig1}.
There, another mass exponential, adapted from the QCD sum rule result for $\Lambda_b$ \cite{BB} and called \lq\lq BB\rq\rq , has also been tested which led to a less pronounced peak of the $\Lambda_c$ DA and the \(p \rightarrow \Lambda_c \) GPDs at $\bar{x}_{10}$. 
When presenting our results we will, for comparison, also show predictions obtained with the BB-type mass exponential \big($\exp[-a_{\Lambda_c(D)} M_{\Lambda_c(D)}(1-\hat{x}')]$\big) for both, $\Lambda_c$ and $D^-$.
The tilde and hat over the arguments in Eqs.~(\ref{eq_pimson_LCWF}) and (\ref{eq_Dmeson_LCWF}) indicate that these definitions of the LCWFs refer to frames in which the corresponding particles move along the 3-direction \cite{overlap}.
Transverse boosts that leave the plus components of four vectors unchanged, lead back to our CMS.
The relationship between momentum fractions and momenta with a tilde to those with a hat is uniquely determined by $\mathbf{\Delta}_\perp$.

With these models for the valence (anti)quark LCWFs of the $\pi^-$ and the $D^-$ we are now able to calculate the $\pi^-\rightarrow D^-$ transition GPD ${H}^{\overline{cu}}_{\pi D}$ and the corresponding form factor $G$ by means of Eqs.~(\ref{eq_mesonic_GPDs}) and (\ref{eq_piDff}), respectively. The analytic expression for ${H}^{\overline{cu}}_{\pi D}$ is given in App.~\ref{app_overlap}.
Results for ${H}^{\overline{cu}}_{\pi D}$ and $G$ are presented in Fig.~\ref{fig_GPDs}. The GPD ${H}^{\overline{cu}}_{\pi D}$ exhibits the expected pronounced peak near $\bar{x}_{20}$, with the peaking value being slightly shifted towards larger values of $\bar{x}_2$ for increasing $\mathbf{\Delta}_\perp^2$ (or $-t^\prime$).
The right plot in Fig.~\ref{fig_GPDs} shows the corresponding form factor as function of $|t^\prime|$ for different values of Mandelstam $s$.
Interestingly it exhibits only a weak dependence on $s$. The results resemble very much those of the $p\rightarrow \Lambda_c$ transition GPDs, discussed in detail in Ref.~\cite{schweig1}.
If we had taken the BB mass exponential for the $D^-$ wave function instead of the KK one the $\pimeson \rightarrow \Dmeson$ transition GPD \( {H}^{\overline{cu}}_{\pi D} \) would become broader and the shift of its maximum to larger $\bar{x}_2$ with increasing $\mathbf{\Delta}_\perp^2$ is somewhat faster than for the KK mass exponential.
For $|t^\prime|\lesssim 3$~GeV$^2$ the BB mass exponential provides a considerably larger transition form factor $G$ than the KK mass exponential.

\section{Observables} \label{sec_results}
The unpolarized differential cross section for $\process$ is (neglecting $m_p$ and $m_\pi$ in the phase-space factor):
\begin{align}
\begin{aligned}
\frac{d\sigma}{d\Omega}&=\frac{1}{64 \pi^2 s}
%\frac
{\sqrt{1-\frac{(M_{\Lambda_c}+M_D)^2}{s}}}
%{\sqrt{1-\frac{(m_p + m_\pi)^2}{s}}}
%\frac
{\sqrt{1-\frac{(M_{\Lambda_c}-M_D)^2}{{s}}}}
%{\sqrt{1-\frac{(m_p - m_\pi)^2}{s}}}
\\
& \phantom{=}\times \left[ \, |\mathcal{M}_{++}|^2+ |\mathcal{M}_{+-}|^2\, \right]\, .
\end{aligned}
\end{align}
The differential cross section predictions for several  values of $s$
%\( s = 20\), $25$ and \(30~\gev \)
are presented in Fig.~\ref{fig_dcs}.
The left plot is the result obtained with our standard parameterization with the KK mass exponential. The right plot shows, for comparison, a calculation with the BB mass exponential.
The forward peak of the cross section is obviously more pronounced for the latter.
The shaded bands take the uncertainties of the $\Lambda_c$ and $D^-$ LCWF parameters into account.
The band corresponds to a variation of $P_{\Lambda_c}$ and $P_D$ between $0.8$ and $1$, of $f_D$ within the experimental uncertainties and of $\langle \mathbf{k}^2_{\perp\, c}\rangle^{1/2}_{\Lambda_c}$ within a range of $417\pm 42$~MeV (see also Refs.~\cite{schweig1} and \cite{gor}) and {from taking $s$ instead of $(\bar{x}_{10} \bar{x}_{20} s)$ as argument of $\alpha_s$.}
\begin{figure}[t]
 \centering
\includegraphics[width=.48\textwidth]{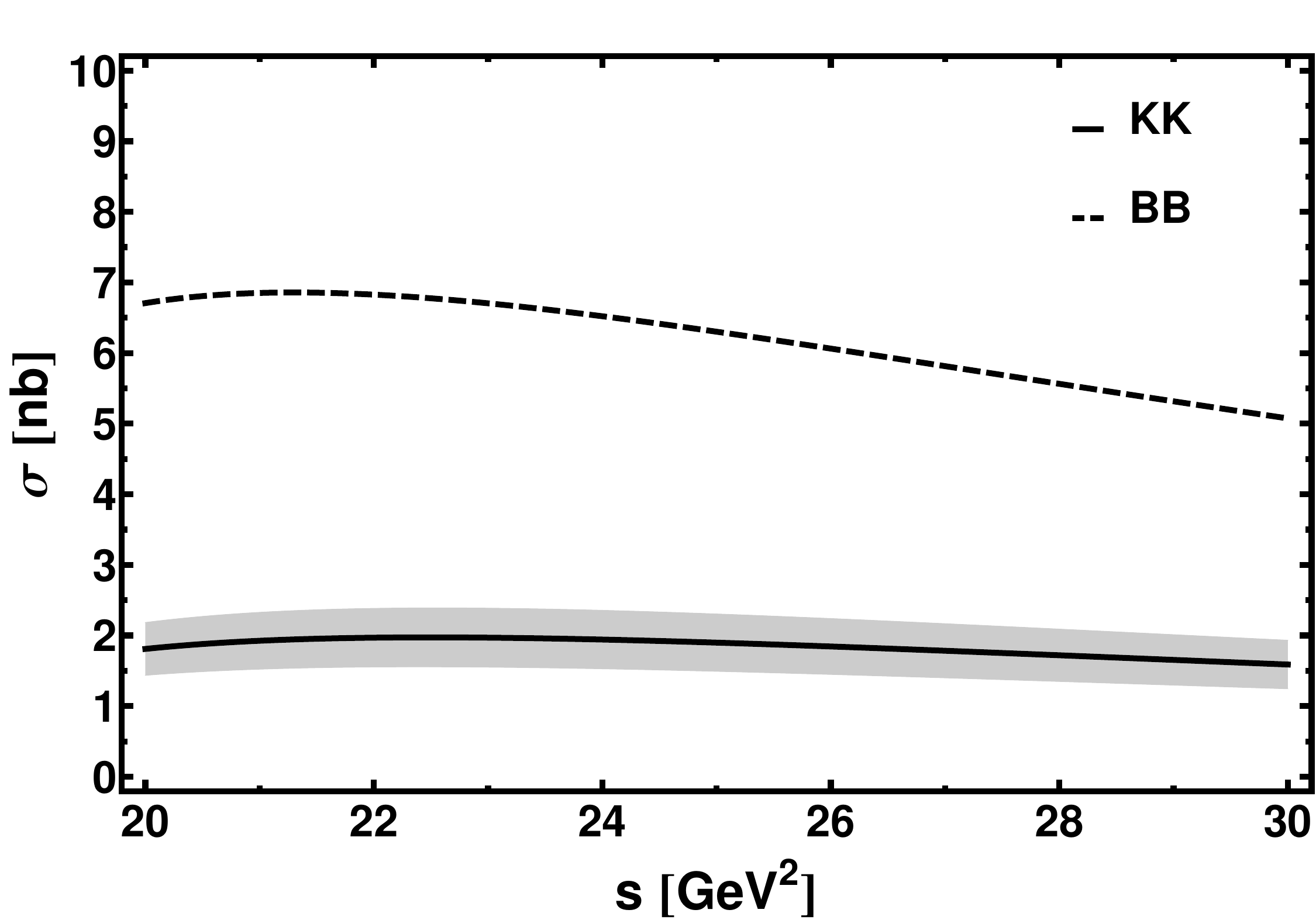}
\caption{Our prediction for the integrated cross section \( \sigma \) versus Mandelstam \( s \) (solid line with error band). For comparison we give also predictions obtained with the BB mass exponential (dashed line, see Ref.~\cite{schweig1}).}
\label{fig_ics}
\end{figure}

The integrated cross section is plotted in Fig.~\ref{fig_ics} for both, the KK and the BB mass exponentials.
As for the differential cross section, we have also made an error assessment in case of the KK mass exponential.
A comparable error band is also found for the BB mass exponential.
The differences between the predictions obtained with different analytic forms of the $\Lambda_c$ and $D^-$ LCWFs are obviously much larger than the variations coming from parametric errors in the wave functions.
The integrated cross sections are of the order of nb with the BB mass exponential giving the larger results.
This is the order of magnitude that has also been found for $ \bar{p} p \rightarrow  \bar{\Lambda}_c^- \Lambda_c^+ $~\cite{schweig1} and $ \bar{p} p \rightarrow \overline{D}^0 D^0$~\cite{gor}, when treated within the generalized parton framework.
It is in accordance with old AGS experiments at  \( s \approx 25~\mbox{GeV}^2 \) which found upper bounds of \( 7~\mbox{nb} \) for \( \pi^-\, p\rightarrow D^{\ast -} \, \Lambda_c^+ \) and \( \approx 15~\mbox{nb} \) for \( \process \)~\cite{Christenson:1985ms}.
A new and more precise measurement of these cross sections would be highly welcome.

For $0+1/2\rightarrow 0+1/2$ processes one has three linearly independent polarization observables, one single-spin observable and two spin correlations. Single-spin observables vanish in lowest order perturbation theory, but our approach provides non-trivial predictions for spin correlations.
We consider the polarization transfers
\be
D_{LL}=D_{SS}=\frac{|\Phi_{++}|^2-|\Phi_{+-}|^2}{|\Phi_{++}|^2+|\Phi_{+-}|^2}\, ,
\ee
\noindent and
\be\label{eq:DZX}
D_{LS}=\frac{2\, \mathrm{Re}(\Phi_{++} \Phi_{+-}^\ast)}{|\Phi_{++}|^2+|\Phi_{+-}|^2}
\ee
as the two independent, nontrivial spin correlations. The labels \lq\lq S\rq\rq\ and \lq\lq L\rq\rq\ denote longitudinal and sideways (in the scattering plane) polarization directions (cf. Ref.~\cite{schweig1}).

\begin{figure*}[t]
 \centering
\includegraphics[width=.49\textwidth]{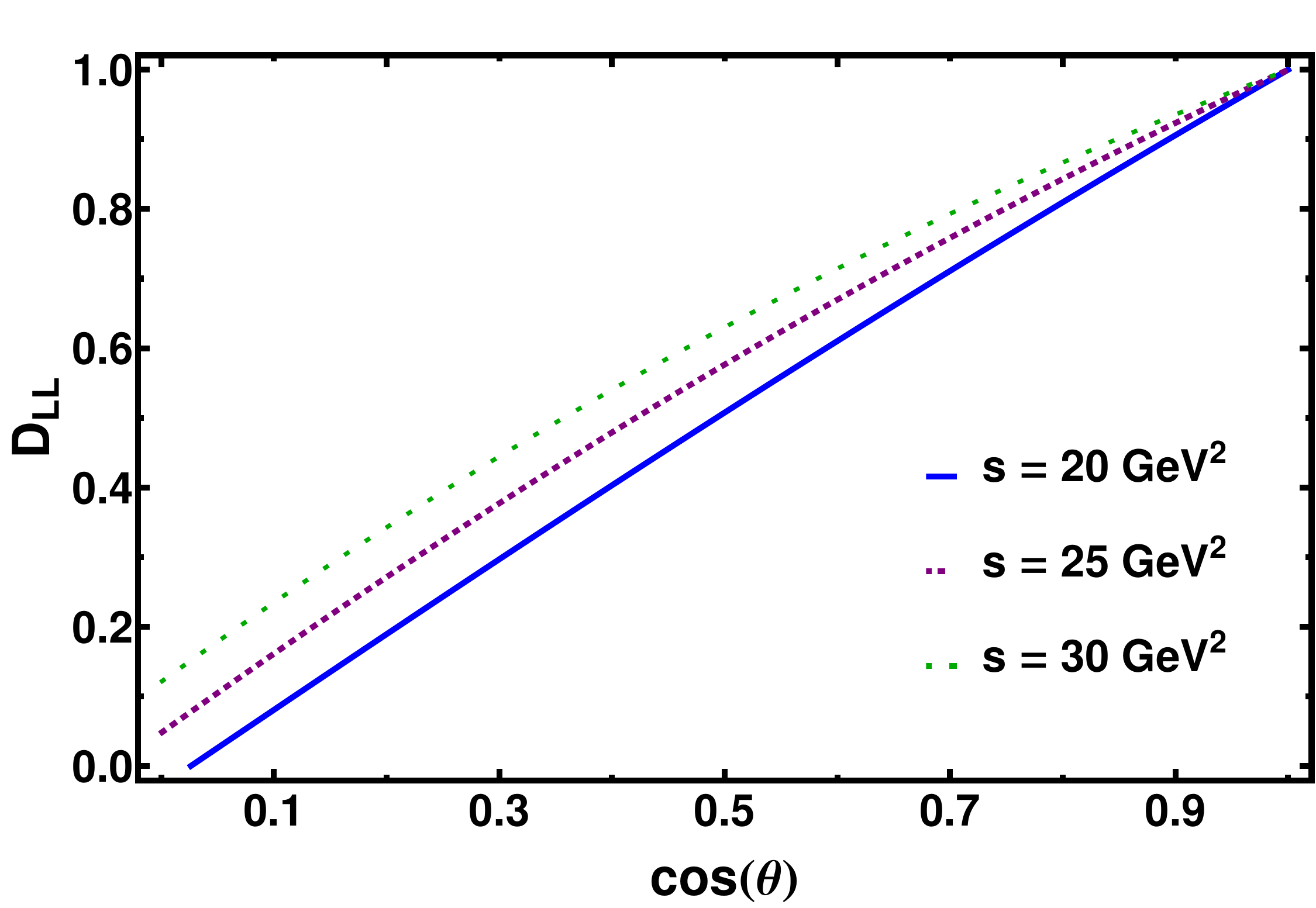}
\includegraphics[width=.49\textwidth]{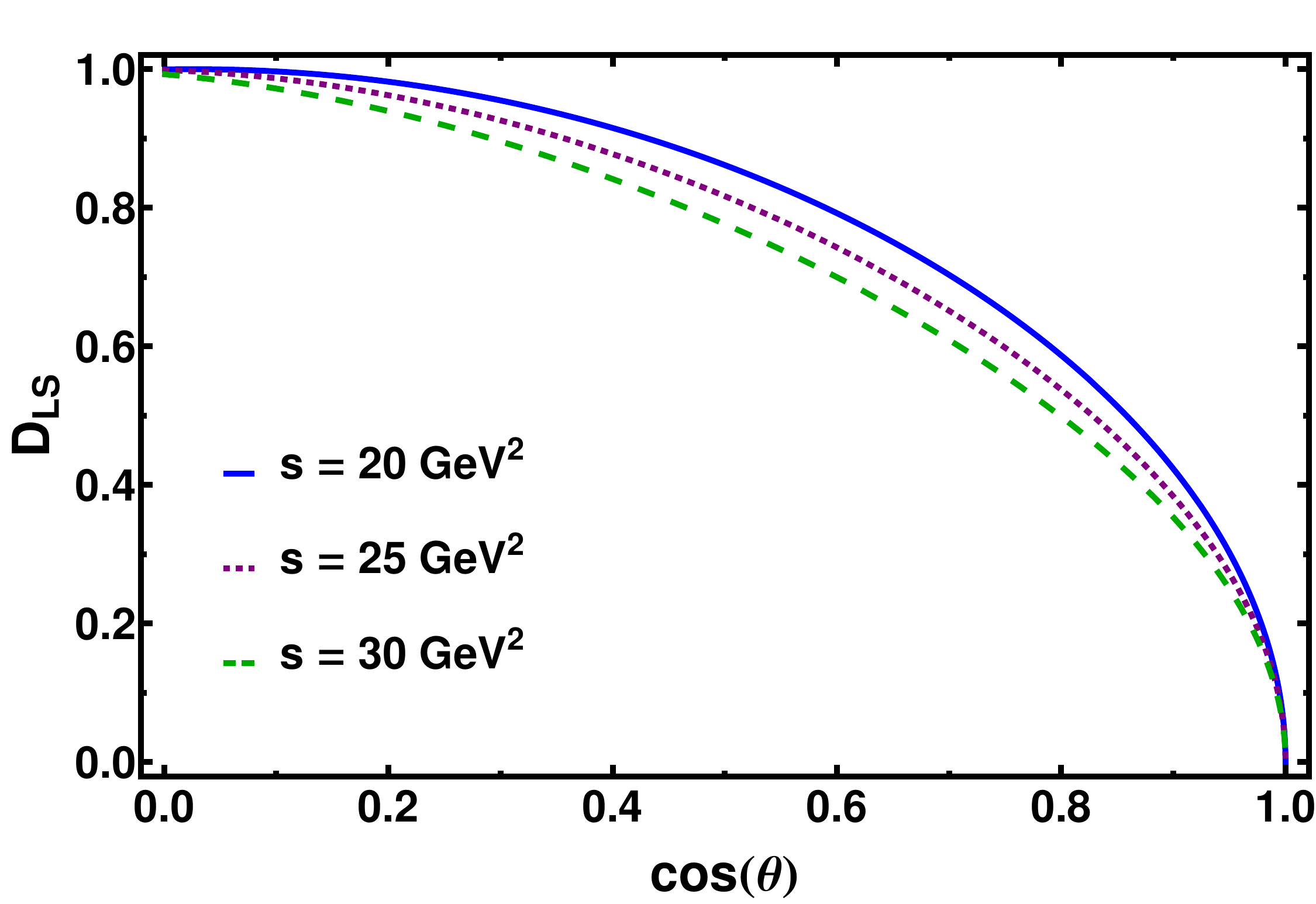}
\caption{
The spin-correlation parameters $D_{XX}$ (left plot) and \( D_{ZX} \) (right plot) versus $\cos \theta$ for $s=20$, $25$ and $30$~GeV$^2$.}
\label{fig_KKspinobs1}
\end{figure*}

The $\Phi_{\tilde\mu^\prime \tilde\mu}$ are CMS helicity amplitudes which are related to our LC helicity amplitudes $\mathcal{M}_{\mu^\prime \mu}$, as defined in Eq.~(\ref{eq_procamp}), by means of an appropriate Melosh rotation \footnote{Like in Ref.~\cite{schweig1} only the Melosh transformation of the $\Lambda_c$ helicity is considered since that of the $p$ plays a minor role.} (see Ref.~\cite{schweig1}). For a reasonable probability of about $10\%$ to find the $c$ quark with helicity opposite to the $\Lambda_c$ helicity in the $\Lambda_c$, the form factors $R_V$ and $S_T$ differ by less then $2\%$~\cite{schweig1}. As a consequence all the form factors and thereby the whole model dependence nearly cancel out in $D_{LL}$ and $D_{LS}$. The energy dependence of $D_{LL}$ and $D_{LS}$ is plotted in Fig.~\ref{fig_KKspinobs1} for the KK mass exponential. It occurs to be very mild {over the considered energy range}. The corresponding plots for the BB mass exponential look more or less the same, which confirms the approximate independence of $D_{LL}$ and $D_{LS}$ on the choice of the GPDs.

\section{Summary and Conclusions}\label{sec_sum}
In this paper we have investigated the exclusive process \( \process \) within the generalized parton picture.
Thereby we have extended foregoing work on $\bar{p} ~p \rightarrow \bar{\Lambda}^-_c ~\mylambda$~\cite{schweig1}, where $p\rightarrow \Lambda_c$ transition GPDs were introduced for the first time.
The analysis of \( \process \) is analogous to the one for $\bar{p} ~p \rightarrow \bar{\Lambda}^-_c ~\mylambda$, the only new ingredients being the $\pi^-\rightarrow D^-$ transition GPDs which replace those for the $\bar{p}\rightarrow \bar{\Lambda}_c$ transition.
Starting with a double-handbag-type mechanism for the production of the charmed hadrons the arguments for factorization into the hard partonic subprocess $\bar{u}\, u\rightarrow \bar{c}\, c$ and soft hadronic matrix elements, which describe the $\pi^-\rightarrow D^-$ and $p\rightarrow \bar{\Lambda}_c$ transitions, are quite the same as for  $\bar{p} ~p \rightarrow \bar{\Lambda}^-_c ~\mylambda$.
Under the assumption that  the transition GPDs are strongly peaked for momentum fractions close to $m_c/M_{\Lambda_c, D}$ the process amplitude simplified further and became just the product of the hard-scattering amplitude with generalized transition form factors, which are kind of moments of the GPDs.
To model the GPDs and make numerical predictions we have employed an overlap representation in terms of LCWFs for the valence Fock states of the hadrons involved.

Interesting for planned experiments, e.g. at J-PARC or at COMPASS, we found the integrated cross section well above production threshold ($s\gtrsim 20$~GeV$^2$) to be of the order of nb, depending on the models for the hadron LCWFs.
Our result is in accordance with experimental evidence on $\pi^-\, p\rightarrow D^{\ast -} \, \Lambda_c^+$~\cite{Christenson:1985ms}.
The size of the $\pi^- p\to D^-\Lambda_c^+$ cross section is typical for the exclusive production of charmed hadrons, like $\bar{p} ~p \rightarrow \bar{\Lambda}^-_c ~\mylambda$~\cite{schweig1}, $\bar{p} ~p \rightarrow D^0~\overline{D}^0$~\cite{gor} and $\gamma\, p\rightarrow \overline{D}^0\, \Lambda_c^+$~\cite{kofler} when treated within the same kind of factorization approach that has been applied here.
We expect a cross section of this size also for the case of a pion-induced production of longitudinally polarized $D^*$ mesons in a straightforward extension of our model.
The calculated spin correlation parameters, on the other hand, were seen to be nearly independent on the models for the LCWFs. This means that those spin correlations are mostly determined by the hard partonic subprocess and may thus give us some clues on how charm is produced on the partonic level.

Exclusive production of charmed hadrons has also been addressed to in Regge models.
For $\pi^-p\to D^-\Lambda_c^+$ one has to consider the $D^*$ trajectory.
Its exchange leads to a characteristic factor
\begin{equation}
\sim \left(\frac{s}{s_0}\right)^{\alpha_{D^*}(t_0)}
\end{equation}
for the forward scattering amplitude.
With a typical trajectory $\alpha_{D^*}(t)\simeq -1 +t/2\,{\rm GeV}^{-2}$ \cite{kaidalov,brisudova} and the still sizable value of $|t_0|$ for $s$ in the range \(20-30 ~\mbox{GeV}^2 \) one notices a strong suppression of the $D\Lambda_c$ channel as compared to the strangeness channel $K\Lambda$, where the $K^*$ trajectory is exchanged. In the strangeness channel $|t_0|$ is very small
for $s\simeq 20 \gev$. Thus, at $t=t_0\simeq 0$ the $K^*$ trajectory takes a value of about \(  0.4 \).
In addition to the strong charm/strange suppression through the different trajectories and values of $t_0$ there is the issue of flavor symmetry breaking in the Regge residues and in the scale parameter, $s_0$.
For the scale parameter it is usually relied on the quark-gluon string model of binary reactions \cite{kaidalov}.
In detail the differences in the Regge parameters and in the residues lead to substantial differences in the results for the charm/strange suppressions.
Thus, in the recent work \cite{Kim:2014qha} a suppression factor of about $10^{-3}$ has been obtained and hence a cross section of the order of nb in agreement with our finding.
In sharp contrast to \cite{Kim:2014qha} Khodjamirian {\it et al} \cite{Khodjamirian:2011sp}
found a much milder charm/strange suppression.
Thus, for instance, for the $\bar{p}p \to\bar{\Lambda}_c^- \Lambda_c^+$ cross section they obtained a value which is about two orders of magnitude larger than the estimate in our partonic picture \cite{schweig1}.
Results for our process, $\pi^-p\to D^-\Lambda_c^+$, are not quoted in \cite{Khodjamirian:2011sp}.
We stress that in our model $SU(4)$-flavor-symmetry breaking (in addition to the one from the hadron and quark masses) occurs due to the flavor dependence of the hadron wave functions which diminishes the $p\rightarrow \Lambda_c$ and $\pi^-\rightarrow D^-$ overlaps considerably as compared to the $p\rightarrow \Lambda$ and $\pi \rightarrow K$ ones \cite{quadder}.

Exclusive charm production near threshold has also been estimated within hadronic models with unreggeized meson exchanges \cite{Haidenbauer:2014rva,Haidenbauer:2009ad}.
The $SU(4)$-symmetry breaking in this approach is hidden in initial and final-state interactions and phenomenologically parameterized vertex form factors.
In the hadronic model the estimated cross sections are about a factor of \( 100 - 1000\) larger than ours.
Cross sections as large as predicted by hadronic or some of the Regge models would also indicate that, in contrast  to our assumption, charm is produced non-perturbatively which means that (non-perturbative) intrinsic charm of the proton must be taken into account.
This could, in principle, be done within our approach, but it is hardly conceivable that the small amount of intrinsic charm in the proton that is compatible with inclusive data~\cite{Jimenez-Delgado:2014zga} could increase the cross section for the
exclusive production of charmed hadrons by two or three orders of magnitude.
Experimental data for processes like $ \process $,  $\bar{p} ~p \rightarrow \bar{\Lambda}^-_c ~\mylambda$, $\gamma\, p\rightarrow \overline{D}^0\, \Lambda_c^+$ and $\bar{p} ~p \rightarrow D^0~\overline{D}^0$ up to several GeV above production threshold would thus be highly desirable to pin down the production mechanism of charmed hadrons and shed some more light on the question of non-perturbative intrinsic charm in the proton.

\section{Acknowledgments}\label{sec_ack}
\noindent We acknowledge helpful discussions with Alexander Goritschnig.
S.K. is supported by the Fonds zur F\"orderung der wissenschaftlichen Forschung in \"Oster\-reich via FWF DK W1203-N16.

\appendix

\section{The $\pi\rightarrow D$ Transition GPD $H_{\pi D}^{\overline{cu}}$}\label{app_overlap}
Employing the $\pi$- and $D$-meson LC wave functions, Eqs.~(\ref{eq_pimson_LCWF}) and (\ref{eq_Dmeson_LCWF}), in Eq.~(\ref{eq_mesonic_GPDs}) the $\pi\rightarrow D$ transition GPD $H_{\pi D}^{\overline{cu}}$ takes on the form:
 \begin{align}\label{eq_Hcu}
 \begin{aligned}
  H_{\pi D}^{\overline{cu}} &= \frac{N_{\pi} N_{D}}{16 \pi^2} \frac{\left( \bar{x}_{2} - 1 \right) \left( \bar{x}_{2}^2 - \eta^2 \right)}{a_{\pi}^2 \left(\bar{x}_{2} - \eta \right) \left(1 + \eta \right)^2+ a_{D}^2 \left(\bar{x}_{2} + \eta \right) \left(\eta - 1 \right)^2} \\
  & \times \exp \left[ \frac{-\mathbf{\Delta}_{\perp}^2 a_{\pi}^2 a_{D}^2 \left(1-\bar{x}_2 \right)} {a_{\pi}^2 \left( \bar{x}_{2} - \eta \right) \left(1 + \eta \right)^2 + a_{D}^2 \left(\bar{x}_{2} + \eta \right) \left( \eta - 1\right)^2} \right] \\
  & \times \exp \left[- f\left( \bar{x}_{2} \right)\right]\, .
 \end{aligned}
 \end{align}
The function \( f \left( \bar{x}_{2} \right) \) depends on the chosen mass exponential in the $D$-meson LC wave function. It is
\be\label{eq_KKme}
f_{KK} \left( \bar{x}_2 \right) = \frac{a_D^2 M_{D}^2 \left(\bar{x}_{2} - \eta + \bar{x}_{20} \left( \eta - 1 \right) \right)^2}{\left(\bar{x}_{2} - \eta \right) \left( 1 -\bar{x}_2 \right)}
\ee
for the KK mass exponential~\cite{KK} and
\be\label{eq_BBme}
f_{BB} \left( \bar{x}_2 \right) =\frac{a_D M_D \left( \bar{x}_{2} - 1\right)}{\eta - 1}
\ee
for the BB~\cite{BB} one. Analytic expressions for the $p\rightarrow \Lambda_c$ transition GPDs can be found in Ref.~\cite{schweig1}.

The transverse momentum transfer and the components of the hadron momenta which are necessary to calculate the skewness parameters (cf. Eqs.~(\ref{eq_skewness}) and (\ref{eq_qbar_eta_def})) are given by:
\be\label{eq_deltaperp}
\mathbf{\Delta}_\perp^2= \frac{s\, \Lambda^2 \Lambda^{\prime 2} \sin^2\theta}{\Lambda^2+\Lambda^{\prime 2}+2 \Lambda \Lambda^\prime \cos\theta}\, ,
\ee
\be
p^+ =\frac{1}{2}\sqrt{\frac{s}{2}}\left(1+\frac{m_p^2-m_\pi^2}{s}+\sqrt{\Lambda^2
-\frac{\mathbf{\Delta}_\perp^2}{s}}\right) ,
\ee
\be
p^{\prime +}=\frac{1}{2}\sqrt{\frac{s}{2}}\left(1+\frac{M_{\Lambda_c}^2-M_D^2}{s}+\sqrt{\Lambda^{\prime 2}-\frac{\mathbf{\Delta}_\perp^2}{s}}\right) ,
\ee

\be
q^-=\frac{1}{2}\sqrt{\frac{s}{2}}\left(1+\frac{m_\pi^2-m_p^2}{s}+\sqrt{\Lambda^2
-\frac{\mathbf{\Delta}_\perp^2}{s}}\right) ,
\ee
\be\label{eq_qpminus}
q^{\prime -}=\frac{1}{2}\sqrt{\frac{s}{2}}\left(1+\frac{M_D^2 - M_{\Lambda_c}^2}{s}+\sqrt{\Lambda^{\prime 2}-\frac{\mathbf{\Delta}_\perp^2}{s}}\right) ,
\ee
with
\be
\Lambda^2=\left(1-\frac{(m_p+m_\pi)^2}{s}\right)\left(1-\frac{(m_p-m_\pi)^2}{s}\right)\, ,
\ee
\be
\Lambda^{\prime 2}=\left(1-\frac{(M_{\Lambda_c}+M_D)^2}{s}\right)\left(1-\frac{(M_{\Lambda_c}+M_D)^2}{s}\right)\, ,
\ee
and
\be
s = \frac{4 |\mathbf{p}|^2}{\Lambda^2}\, ,
\ee
where $\mathbf{p}$ is the proton momentum and $\theta$ the scattering angle in the CM system.

\section{Partonic Scattering Amplitudes}\label{app_part}
In this appendix we list the independent partonic subprocess amplitudes necessary to calculate the hadronic LC-helicity amplitudes in the peaking approximation. As mentioned in the text these amplitudes can be written in terms of pure hadronic quantities for which we take approximate expressions that are obtained from Eqs.~(\ref{eq_deltaperp})-(\ref{eq_qpminus}) above by setting $m_p=m_\pi=0$ and an average heavy-hadron mass $M=\sqrt{M_{\Lambda_c} M_D}$:
\begin{eqnarray}
H_{+-,+-}&=& \frac{4 \pi \alpha_s(\bar{x}_{10} \bar{x}_{20} s)}{s} \frac{4(p^+ p^{\prime +}-\mathbf{\Delta}_\perp^2/8)^2+2 M^2 p^{+2}}{p^+ p^{\prime +}}\, ,\nonumber\\
H_{+-,-+}&=& -\frac{4 \pi \alpha_s(\bar{x}_{10} \bar{x}_{20} s)}{s} \frac{\mathbf{\Delta}_\perp^2\left[ 2 (p^+ + p^{\prime +})^2+M^2\right]}{4 p^+ p^{\prime +}}\, ,\nonumber\\
H_{++,-+}&=& \frac{4 \pi \alpha_s(\bar{x}_{10} \bar{x}_{20} s)}{s} \frac{M |\mathbf{\Delta}_\perp|\left[ 2 p^+ p^{\prime +}-\mathbf{\Delta}_\perp^2/8+p^{+2}\right]}{p^+ p^{\prime +}}\, .\nonumber\\
\end{eqnarray}
For the strong coupling constant $\alpha_s$ we take the one-loop expression with four flavors and $\Lambda_{QCD}=0.24$~GeV.
\phantom{the CM system, respectively. For the strong coupling constant $\alpha_s$ we take the one-loop expression with four flavors and $\Lambda_{QCD}=0.24$~GeV.}


\begin{thebibliography}{}

\bibitem{schweig1}
 A.~T.~Goritschnig, P.~Kroll and W.~Schweiger,
Eur.\ Phys.\ J.\ {\bf{A42}}, 43 (2009).

\bibitem{gor}
A.~T.~Goritschnig, B.~Pire and W.~Schweiger,
Phys.\ Rev.\ D\ {\bf{87}}, 014017 (2013); Erratum: Phys. Rev. D {\bf 88}, 079903 (2013).

\bibitem{kofler}
 A.~Goritschnig, S.~Kofler and W.~Schweiger,
 PoS (Photon 2013), 061 (2014).


 \bibitem{Brodsky:LightCone}
S.~J.~Brodsky, H-C.~Pauli and S.~S.~Pinsky,
Phys.\ Rept.\ {\bf{301}}, 299 (1998).

\bibitem{kniehl-kramer}
B.~A.~Kniehl and G.~Kramer,
  %``Charmed-hadron fragmentation functions from CERN LEP1 revisited,''
  Phys.\ Rev.\ D {\bf 74}, 037502 (2006).

\bibitem{KK}
J.~G.~K\"orner and P.~Kroll,
  %``Heavy quark symmetry at large recoil,''
  Phys.\ Lett.\ B {\bf 293}, 201 (1992).

\bibitem{BB}
P.~Ball, V.~M.~Braun and E.~Gardi,
  %``Distribution Amplitudes of the Lambda(b) Baryon in QCD,''
  Phys.\ Lett.\ B {\bf 665}, 197 (2008).


\bibitem{wideangle}
M.~Diehl, T.~Feldmann, R.~Jakob and P.~Kroll,
Eur.\ Phys.\ J.\ {\bf{C8}}, 409 (1999).

\bibitem{Brommel}
D.~Brommel,
DESY-THESIS-2007-023 (2007).

\bibitem{IW1991}
N.~Isgur and M.~B.~Wise,
Nucl.\ Phys.\ B {\bf{348}}, 276 (1991).

\bibitem{overlap}
M.~Diehl, T.~Feldmann, R.~Jakob and P.~Kroll,
Nucl.\ Phys.\ B {\bf{596}}, 33 (2001).

\bibitem{Brodsky:2000xy}
  S.~J.~Brodsky, M.~Diehl and D.~S.~Hwang,
  Nucl.\ Phys.\ B {\bf 596}, 99 (2001).


\bibitem{Bolz:1996sw}
J.~Bolz and P.~Kroll,
Z.\ Phys.\ A {\bf{356}}, 327 (1996).

\bibitem{Feldmann}
T.~Feldmann and P.~Kroll,
Eur.\ Phys.\ J.\ {\bf{C12}}, 99 (2000).

\bibitem{PDG}
J.~Beringer~et~al.,
Phys.\ Rev.\ D {\bf{86}}, 01001 (2012).

\bibitem{Christenson:1985ms}
 J.~H.~Christenson, E.~Hummel, G.~A.~Kreiter, J.~Sculli  and P.~Yamin,
  Phys.\ Rev.\ Lett.\  {\bf 55}, 154 (1985).

\bibitem{Bourrely}
C.~Bourrely, J.~Soffer and E.~Leader,
Phys.\ Rept.\ {\bf{59}}, 95 (1980).

\bibitem{kaidalov}
A.~B.~Kaidalov and P.~E.~Volkovitsky,
  %``Binary reactions in anti-p p collisions at intermediate-energies,''
  Z.\ Phys.\ C {\bf 63}, 517 (1994).

\bibitem{brisudova}
M.~M.~Brisudova, L.~Burakovsky and J.~T.~Goldman,
  %``Effective functional form of Regge trajectories,''
  Phys.\ Rev.\ D {\bf 61}, 054013 (2000).

\bibitem{Kim:2014qha}
S.~H.~Kim, A.~Hosaka, H.~C.~Kim, H.~Noumi and K.~Shirotori, Prog. Theor. Exp. Phys., 103D01 (2014).

\bibitem{Khodjamirian:2011sp}
  A.~Khodjamirian, C.~Klein, T.~Mannel and Y.~M.~Wang,
  Eur.\ Phys.\ J.\ A {\bf 48}, 31 (2012).

\bibitem{quadder}
P.~Kroll, B.~Quadder and W.~Schweiger,
  %``Exclusive Production of Heavy Flavors in Proton - Anti-proton Annihilation,''
  Nucl.\ Phys.\ B {\bf 316}, 373 (1989).

\bibitem{Haidenbauer:2014rva}
J.~Haidenbauer and G.~Krein,
   Phys.\ Rev.\ D {\bf 89}, 114003 (2014).

 \bibitem{Haidenbauer:2009ad}
 J.~Haidenbauer and G.~Krein,
  Phys.\ Lett.\ B {\bf 687}, 314 (2010).


\bibitem{Jimenez-Delgado:2014zga}
  P.~Jimenez-Delgado, T.~J.~ Hobbs, J.~T.~Londergan and W.~Melnitchouk,
  arXiv:1408.1708 [hep-ph].


\end{thebibliography}
\end{document}